\documentclass[9pt,conference]{IEEEtran}

\usepackage{waspaa25} 

\usepackage{bm} 

\usepackage{algorithm}
\usepackage{algpseudocode}
\usepackage[subtle]{savetrees}

\usepackage{siunitx}
\usepackage{soul}
\usepackage{booktabs}
\usepackage{siunitx}
\usepackage{amssymb}
\usepackage{multirow}
\usepackage{makecell}
\usepackage{pifont}
\newcommand{\cmark}{\text{\ding{51}}}
\newcommand{\xmark}{\text{\ding{55}}}

\usepackage{caption}
\title{Unsupervised Multi-channel Speech Dereverberation via Diffusion}

\name{Yulun Wu$^{1}$,
      Zhongweiyang Xu$^{1}$,
      Jianchong Chen$^{1}$,
      Zhong-Qiu Wang$^{2}$,
      Romit Roy Choudhury$^{1}$}
\address{$^{1}$Department of Electrical and Computer Engineering, University of Illinois Urbana-Champaign, Champaign, USA \; \\
$^{2}$Department of Computer Science and Engineering, Southern University of Science and Technology, Shenzhen, China
}

\begin{document}
\maketitle

\begin{abstract}
We consider the problem of multi-channel single-speaker blind dereverberation, where multi-channel mixtures are used to recover the clean anechoic speech. To solve this problem, we propose USD-DPS, {U}nsupervised {S}peech {D}ereverberation via {D}iffusion {P}osterior {S}ampling. USD-DPS uses an unconditional clean speech diffusion model as a strong prior to solve the problem by posterior sampling. At each diffusion sampling step, we estimate all microphone channels' room impulse responses (RIRs), which are further used to enforce a multi-channel mixture consistency constraint for diffusion guidance. For multi-channel RIR estimation, we estimate reference-channel RIR by optimizing RIR parameters of a sub-band RIR signal model, with the Adam optimizer. We estimate non-reference channels' RIRs analytically using forward convolutive prediction (FCP). We found that this combination provides a good balance between sampling efficiency and RIR prior modeling, which shows superior performance among unsupervised dereverberation approaches. An audio
demo page is provided in {\href{https://usddps.github.io/USDDPS_demo/}{https://usddps.github.io/USDDPS\_demo/}}.
\end{abstract}

\section{Introduction}
\label{sec:intro}
In reverberant enclosures like a room, clean speech signals get reflected by walls and are then recorded by microphones, which creates reverberant effects on the recorded speech~\cite{reverberation}. This reverberation effect degrades both the perceptual quality and intelligibility of the clean speech signal. This paper focuses on the problem of recovering the clean speech given the microphone-array recorded reverberant mixtures (the mixture of clean source, echoes, and late reverberations).

Supervised learning with neural networks is one popular approach to solve this problem~\cite{sup1}. It needs a large-scale, synthesized, clean, and reverberant paired dataset, where the neural networks take reverberant mixtures as inputs and use the anechoic clean speech as training labels. Although these methods work well, they are often treated as black-box models, which lack explainability and generalizability.

For unsupervised methods, weighted predicted error (WPE)~\cite{wpe1, wpe2,Yoshioka2012WPESPM} proposes to use delayed linear prediction to estimate late reverberation, which can then be subtracted from the mixture for dereverberation. Despite its simplicity and effectiveness, it does not fully exploit speech and RIR's prior.

Recently, USDnet~\cite{usdnet} has been proposed as an unsupervised neural speech dereverberation method, which allows training a strong dereverberation deep learning model only on reverberant mixtures. Without any clean source as training labels, following UNSSOR~\cite{unssor}, USDnet designs a mixture-constraint loss. 
This loss tries to ensure that the model's output, which is the clean source estimate, can be linearly filtered to reconstruct the multi-channel reverberant mixtures. 
To obtain linear filters during training, forward convolutive prediction (FCP)~\cite{fcp1} is used to estimate filters that maximally align the model output and the reverberant mixture. 
However, since USDnet does not use a prior for the source speech and the RIRs, the dereverberation performance is limited.
To address this problem, we propose to introduce a clean speech diffusion prior and also use an RIR model when estimating the filters.

To introduce a strong prior, diffusion models~\cite{ddpm, score, edm} are popular generative models that can be used to solve inverse problems~\cite{score, dps, pigdm}. \cite{saito2022unsupervisedvocaldereverberationdiffusionbased} uses a diffusion model to refine the dereverberation filter coefficients adaptively. Diffusion posterior sampling (DPS)~\cite{dps} proposes to compute the posterior score by adding a likelihood guidance term to the original prior score. The posterior score can be used for posterior sampling to solve the inverse problem, which has later been extended to audio inverse problems~\cite{cqt} and RIR-informed speech dereverberation~\cite{informed}. However, DPS assumes that the forward operator of the inverse problem is known, which is not the case for blind inverse problems like blind dereverberation/deconvolution. 
BlindDPS~\cite{blinddps} and Fast Diffusion EM~\cite{fastdiffusionem} extend DPS to unknown operators by incorporating a dedicated module to estimate the operator. Similarly, recent studies have worked on a few specific blind inverse problems in the audio domain~\cite{blindbwe, blindhrtf, blindnonlinear, buddy, xu2025arraydpsunsupervisedblindspeech}.

To solve the problem of blind speech dereverberation, BUDDy~\cite{buddy} proposes to estimate the RIR at each DPS sampling step. It uses a parameterized sub-band RIR model with exponential decay for each frequency band. At each DPS sampling step, BUDDy estimates the RIR model parameters iteratively using the Adam optimizer. Then the estimated RIR can be used as the forward operator. However, BUDDy only considers single-channel blind dereverberation.

To solve multi-channel blind speech dereverberation, one obvious method is to extend BUDDy to multi-channel, where each channel has an RIR model for optimization. We call this multi-channel BUDDy. We found that this naive extension is slow as the number of channels increases, as each channel's RIR needs to be estimated by Adam in each DPS step. Thus, we propose USD-DPS, which only estimates the reference channel's RIR using the parameterized sub-band RIR model, and estimates all other channels' RIRs using forward convolutive prediction (FCP), which has an analytical solution. We find that USD-DPS performs much faster than multi-channel BUDDy and improves dereverberation, resulting in state-of-the-art results in multi-channel unsupervised blind speech dereverberation.

\section{Background and Problem Formulation}
\label{sec:background}
In reverberant scenarios with a single speaker and $C$ microphones, define $x_1$ as the target anechoic signal captured at the reference microphone $1$. Then the $C$-channel reverberant mixtures are $y=\{y_c | y_c = h_c * x_1 + n_c, 1\leq c\leq C\}$. Let $h_c$ denote the relative RIR from $x_1$ to the $c^{\text{th}}$ microphone, and let $n_c$ denote the $c^{\text{th}}$ microphone's measurement noise. The task of dereverberation is to recover $x_1$ given $y$.
For future reference, we denote $y_1$ as the reference-channel reverberant mixture, denote $y_{2:C}=\{y_c, 2\leq c \leq C\}$ as non-reference channel mixtures, and denote $h$ as $\{h_1, h_2, ..., h_C\}$.

\subsection{Diffusion Posterior Sampling for Dereverberation}\label{sec:dps}
{\bf Score-based Diffusion: }
Score-based diffusion models~\cite{ddpm, score, edm} are proposed to generate samples from a data distribution $p_{\text{data}}(x_1)$. $x_1$ is reference-channel clean anechoic speech in our case. These models define a diffusion process that transforms the data distribution to a Gaussian, and then a score-matching or denoising objective is used to learn how to reverse the diffusion process from Gaussian noise. Then, during sampling, a noise is first sampled $x_1^{\tau_{\text{max}}}\sim \mathcal{N}(0, \sigma^2(\tau_{\text{max}})I)$, and then gradually transformed to a data sample $x^0_1\sim~p_{\text{data}}(x_1)$ following the probabilistic flow ordinary differential equation as in EDM~\cite{edm}:
\begin{equation}\label{eq:ode}
d{x}_1^\tau = -\sigma(\tau) \nabla_{x_1^\tau} \log p(x_1^\tau) d\tau    
\end{equation}
where we use $\sigma(\tau)=\tau$ and $\nabla_{x_1^\tau} \log p(x_1^\tau)$ is approximated by the score model $s_\theta(x_1^\tau, \sigma(\tau))$ trained with score matching.

{\bf Diffusion Posterior Sampling: }
Since our goal is to recover the reference-channel anechoic clean speech $x_1$ given the multi-channel mixture $y$, we want to sample from the posterior $p(x_1|y)$ using the following probabilistic flow ODE:
\begin{equation}\label{eq:posterior_ode}
d{x}_1^\tau = -\sigma(\tau) \nabla_{x_1^\tau} \log p(x_1^\tau|y) d\tau    
\end{equation}
Following DPS~\cite{dps}, the posterior score is decomposed as:
\begin{equation}\label{eq:bayesian}
    \nabla_{x_1^\tau} \log p(x_1^\tau|y) = \nabla_{x_1^\tau} \log p(x_1^\tau) + \nabla_{x_1^\tau} \log p(y|x_1^\tau)
\end{equation}
where $\nabla_{x_1^\tau}\log p(x_1^\tau)$ is the prior score approximated by the trained score model $s_\theta(x_1^\tau, \sigma(\tau))$, and $\nabla_{x_1^\tau} \log p(y|x_1^\tau)$ is the likelihood score that needs to be approximated. When the RIRs $h_{1:C}$ are known, similar to~\cite{informed}, $\nabla_{x_1^\tau} \log p(y|x_1^\tau)$ can then be approximated using:
\begin{equation}\label{eq:likelihood_aprox}
    \nabla_{x_1^\tau} \log p(y|x_1^\tau) = \nabla_{x_1^\tau} \log p(y|x_1^\tau, h)\simeq~\nabla_{x_1^\tau} \log p(y|\hat{x}_1^0, h)
\end{equation}
where, following the Tweedie's formulae, $\hat{x}_1^0=x_1^\tau-\sigma^2(\tau)s_\theta(x_1^\tau, \tau)$, and, following BUDDy's compressive domain likelihood, $p(y|\hat{x}_1^0, h)=\prod_{c=1}^C \mathcal{N}\left(S_{\text{comp}}({y});\, S_{\text{comp}}(\hat{{x}}_1^0 * {h_c}),\, \eta^2 {I}\right)$.
The STFT domain compression function is defined as $S_{\text{comp}}(\mathbf{y}) = |\text{STFT}(\mathbf{y})|^{2/3} \exp\{j\angle\text{STFT}(\mathbf{y})\}$ with $j$ denoting the imaginary unit, and $\eta$ is the measurement noise level in the compressed STFT domain.

\subsection{BUDDy and RIR model}\label{sec:buddy}
Since the RIR $h$ is usually unknown, BUDDy proposes a signal reverberation model that allows single-channel RIR estimation (assume $h=h_1,y=y_1$ in this subsection). The signal reverberation model $\mathcal{A}_\psi: \mathbb{R}^L \rightarrow \mathbb{R}^L$ models RIR convolution in the STFT domain using subband convolution ($\mathcal{A}_\psi$ includes STFT, RIR sub-band convolution, and iSTFT sequentially; $L$ denotes input and output sample length). Let $H = \text{STFT}(h) \in \mathbb{C}^{N_h \times K}$ denote the STFT of the RIR with $N_h$ time frames and $K$ frequency
bins, and $X$, $Y$ be the STFTs of the clean and reverberant speech. Each frequency band $k$ undergoes independent 1D convolution:
\begin{equation}\label{eq:conv}
Y_{m,k} = H_{m,k} * X_{m,k} = \sum_{n=0}^{N_h-1} H_{n,k} X_{m-n,k}.
\end{equation}
To enable estimation of RIR, $H$ is parameterized with structured priors. $H$'s magnitude response $A \in \mathbb{R}^{N_h \times K}$ is interpolated from a coarsely sampled exponential decay model:
\begin{equation}
A'_{n,b} = w_b e^{-\alpha_b n}, \quad A = \exp\big(\text{lerp}(\log A')\big),
\end{equation}
where $w_b$ and $\alpha_b$ are the weight and decay rate of sub-band $b\in[1, B]$. $H$'s phase $\Phi \in \mathbb{R}^{N_h \times K}$ is directly optimized. The RIR parameters $\psi = \{\Phi, (w_b, \alpha_b)_{b=1}^B\}$ can then be estimated at each sampling step to enable DPS as mentioned in Sec.~\ref{sec:dps}. The estimation objective follows:
\begin{equation}
\hat{\psi}=\arg\min_{\psi} \|S_{\text{comp}}(y)-S_{\text{comp}}(\mathcal{A}_\psi(\hat{x}_1^0))\|_2^2 + R(\psi)
\end{equation}
where $\hat{x}_1^0$ is the current diffusion step's clean speech estimate and $R(\psi)$ is a parameter regularizer. More details can be found in BUDDy~\cite{buddy}.
This estimation objective is optimized using the Adam optimizer with $N_\text{its}$ iterations. After each optimization step update, a projection step ensures STFT consistency and minimum-phase via:
\begin{equation}\label{eq:project}
H \leftarrow \text{STFT}\big(\delta \oplus P_{\min}(\text{iSTFT}(H))\big)
\end{equation}
where $P_{\min}$ is a transform to guarantee $H$ is a minimum-phase system and $\delta \oplus (\cdot)$ replaces the first sample of the RIR to be 1, making sure that the direct path happens at the first sample.

Then at each DPS step, after $\hat{\psi}$ is estimated as above using the current $\hat{x}_1^0$, the likelihood score approximation follows:
\begin{equation}\label{eq:buddy_likelihood_aprox}
    \nabla_{x_1^\tau} \log p(y|x_1^\tau) \simeq~\zeta(\tau)\nabla_{x_1^\tau}\left\|S_{\text{comp}}({y})-S_{\text{comp}}(A_{\hat{\psi}}(\hat{x}_1^0))\right\|_2^2
\end{equation}
$\zeta(\tau)$ is the likelihood guidance parameter usually set empirically. This likelihood score can also be viewed as a step towards a mixture consistency constraint, that the RIR filtered output is close to the reverberant mixture.

\vspace{-5pt}
\section{Method}
\subsection{Likelihood Score Approximation}
\label{sec:likelihood}
 As discussed in Sec.~\ref{sec:dps}, to use
 DPS for multi-channel blind dereverberation, we need an approximation of the likelihood score $\nabla_{x_1^\tau} \log p(y|x_1^\tau)$. However, we cannot use the approximation in DPS because in Eq.~\ref{eq:likelihood_aprox}, the multi-channel RIRs $h$ are unknown. Thus, at each DPS step, we have to estimate all-channel RIRs first and then use the estimated RIRs. The most straightforward way is to extend BUDDy to multi-channel, which we call MC-BUDDy.

{\bf MC-BUDDy}: To extend BUDDy to be multi-channel, we instantiate an RIR model (mentioned in Sec.~\ref{sec:buddy}) for each microphone channel, resulting in $C$ RIR models to be optimized for $N_\text{its}$ iterations at each DPS step. One slight modification is the projection step as in Eq.~\ref{eq:project}, where $\delta\oplus (\cdot)$ replaces the first sample of the RIR to be 1. Since different channels' direct paths should have a time delay, MC-BUDDy only applies this operation for the reference channel (channel 1). For all other channels, the projection step is simply $H \leftarrow \text{STFT}(P_{\min}(\text{iSTFT}(H)))$. Then, similar to BUDDy, MC-BUDDy estimates $C$-channel RIRs' parameters $\{\hat{\psi}_1, \hat{\psi}_2, ..., \hat{\psi}_C\}$ at each DPS step, and then these estimated RIRs are used to approximate the likelihood score following:
{\small
\setlength{\abovedisplayskip}{5pt} 
\setlength{\belowdisplayskip}{5pt} 
\begin{equation}\label{eq:mcbuddy_likelihood_aprox}
    \nabla_{x_1^\tau} \log p(y|x_1^\tau) \simeq~\zeta(\tau)\nabla_{x_1^\tau}\sum_{c=1}^{C}\left\|S_{\text{comp}}(y_c)-S_{\text{comp}}(A_{\hat{\psi}_c}(\hat{x}_1^0))\right\|_2^2
\end{equation}
}
However, one drawback of MC-BUDDy is that estimating all $C$ channel RIRs at every DPS step is computationally slow, and scales up as $C$ increases. Intuitively, all channels' RIRs should be similar in decaying time like RT60, as they are in the same acoustic environment. Thus, we propose USD-DPS, which only uses one RIR model for the reference channel.

{\bf USD-DPS}: As modeling all channels' RIRs is computationally slow, we propose to only use the RIR model mentioned in Sec.~\ref{sec:buddy} for the reference channel. However, this only allows us to estimate the likelihood of the reference-channel mixture. Thus, we propose to estimate all other non-reference channels' RIRs using forward convolutive prediction (FCP)~\cite{fcp1, fcp2}, without using any RIR models. FCP formulates filter estimation as a weighted least squares problem, which has an analytical solution, so it is computationally fast.

Similar to BUDDy's RIR model, FCP also models reverberation as sub-band convolution in the STFT domain as in Eq.~\ref{eq:conv}. Let $Y^c=\text{STFT}(y_c)$, $H^c=\text{STFT}(h_c)$, for $1\leq c\leq C$. Given an STFT domain source estimate $\hat{X}$ and $\{Y^1, Y^2, ..., Y^C\}$, FCP estimates the $c^\text{th}$ channel RIR $H^c$ by solving the following minimization problem:
{\small
\vspace{-5pt}
\setlength{\abovedisplayskip}{5pt} 
\setlength{\belowdisplayskip}{5pt} 
\begin{align}
    &\hat{H}^c = \text{FCP} (Y^c, \hat{X})=  \underset{{H}^c}{\arg\min} \sum_{m,k} \frac{1}{\hat{\lambda}^c_{m,k}} \left|Y^c_{m,k} - \sum_{n=0}^{N_h'-1} H^c_{n,k} \hat{X}_{m-n,k}\right|^2\label{eq:fcp_obj}\\
    &\hat{\lambda}^c_{m,k} = \frac{1}{C} \sum_{c=1}^C |Y^{c}_{m, k}|^2+ \epsilon\cdot \max\limits_{m,k}\frac{1}{C} \sum_{c=1}^C |Y^{c}_{m,k}|^2\label{eq:fcp_lambda}
\end{align}
}
As shown above, FCP is solving a weighted least squares problem, so it has an analytical solution as in~\cite{fcp1, fcp2}. $N_h'$ is the number of frames of the RIR filter. The weight $\hat{\lambda}^c_{m,k}$ aims to prevent the estimated RIR from overfitting to high-energy STFT bins, and $\epsilon$ is a tunable hyperparameter to adjust the weight. 

In this context, in USD-UPS we propose to use FCP to estimate non-reference channel RIRs $\hat{H}^2, \hat{H}^3, ..., \hat{H}^C$. Similar to $\mathcal{A}_{\psi}$, we define the operator $\mathcal{A}_{\hat{H}^c}(\cdot): \mathbb{R}^L \rightarrow \mathbb{R}^L$ as the sequential operation of STFT, sub-band convolution with $\hat{H}^c$, and iSTFT. Then the reference-channel RIR estimated by the RIR model and the non-reference channels RIRs estimated by FCP lead to the likelihood approximation below:
{\small
\setlength{\abovedisplayskip}{5pt} 
\setlength{\belowdisplayskip}{5pt} 
\begin{align}\label{eq:usddps_likelihood_aprox}
    \nabla_{x_1^\tau} \log &p(y|x_1^\tau) \simeq~\zeta(\tau)\nabla_{x_1^\tau}\Bigg(\left\|S_{\text{comp}}({y_1})-S_{\text{comp}}(A_{\hat{\psi_1}}(\hat{x}_1^0))\right\|_2^2 \nonumber\\
    &+ \lambda'\sum_{c=2}^C \Big\|S_{\text{comp}}({y_c})-S_{\text{comp}}(A_{\hat{H}^c}(\hat{x}_1^0))\Big\|_2^2\Bigg)
\end{align}
}
where $\lambda'$ is a hyperparameter to adjust the non-reference channels' likelihood guidance. Besides computational efficiency, the FCP filter estimation process is differentiable~\cite{fcp1}, which provides more direct guidance for the likelihood score~\cite{usdnet, unssor}.

\subsection{Algorithm}
Our USD-DPS dereverberation algorithm is shown in Algorithm~\ref{alg:inference}, which uses the probabilistic flow ODE shown in Eq.~\ref{eq:posterior_ode} while using the likelihood approximation in Eq.~\ref {eq:usddps_likelihood_aprox}. For the ODE sampling discretization with $N$ discretized steps, we use the same scheduler and sampler as BUDDy~\cite{buddy}. Thus, in Algorithm~\ref{alg:inference}, $x_1^n=x_1^{\tau(n)}$, where $\tau(n)$ is the diffusion time at discretized step $n$, following the sampling scheduler. As in lines 1-2 of Algorithm~\ref{alg:inference},  WPE~\cite{wpe1} is first applied to process the input mixtures, and the output is used to initialize the DPS initial point. Then, starting from line 4, $N$ DPS steps are used to gradually denoise the WPE-initialized $x_1^N$ to the clean speech $x_1^0$. Lines 5-6 first get the current score using the pre-trained score network, and then the current clean speech estimate $\hat{x}_1^0$ is derived using Tweedie's formula. Then $\hat{x}_1^0$ is rescaled to an empirical standard deviation of $0.05$. Lines 8-14 use the RIR model to estimate reference-channel RIR parameters, similar to BUDDy. Then, lines 15-19 use FCP to estimate all non-reference channels' RIRs. With all-channel RIRs estimated, lines 20 and 21 get the log likelihood of the reference channel mixture and the non-reference channel mixtures (mixture consistency constraint), respectively. Line 22 then calculates the likelihood score, which is further used to update a DPS step as in line 23.

\begin{algorithm}[t]
\caption{USD-DPS}
\label{alg:inference}
\begin{algorithmic}[1]
\Require multi-channel reverberant speech ${y}$
\State ${x}_{\text{init}} \gets \text{WPE}({y})$ \Comment{\textcolor{magenta}{WPE Warm initialization}}
\State Sample ${x}_N \sim \mathcal{N}({x}_{\text{init}}, \sigma_N^2 {I})$
\State Initialize $\psi_1$ \Comment{\textcolor{magenta}{Initialize ref-channel RIR parameters}}

\For{$n = N, \dots, 1$}
    \State $x_1^n \gets s_\theta({x}_1^n, \tau_n)$ \Comment{\textcolor{magenta}{Evaluate score model}}
    \State $\hat{{x}}_1^n \gets {x}_1^n - \sigma_n^2 s_n$ \Comment{\textcolor{magenta}{Get one-step denoising estimate}}
    \State $\hat{{x}}_1^0 \gets \text{Rescale}(\hat{{x}}_1^0)$
    \State $\psi_1^0 \gets \psi_1$ \Comment{\textcolor{magenta}{Use the RIR parameters from last step}}

    \For{$j = 0, \dots, N_{\text{its}}$} \Comment{\textcolor{magenta}{Ref-channel RIR estimation}}
        \State $\mathcal{J}_{\text{RIR}}(\psi_1^j) \gets \|(S_{\text{comp}}(y_1)-S_{\text{comp}}(\mathcal{A}_{\psi_1^j}(\hat{x}_1^0))\|_2^2 + R(\psi_1)$
        \State $\psi_{1}^{j+1} \gets \psi_{1}^j - \text{Adam}(\nabla \mathcal{J}_{\text{RIR}}(\psi_{1}^j))$ \Comment{\textcolor{magenta}{Optim. step}}
        \State $\psi_{1}^{j+1} \gets \text{project}(\psi_{1}^{j+1})$ \Comment{\textcolor{magenta}{Projection step}}
    \EndFor

    \State $\psi_{1} \gets \psi_{1}^{N_\text{its}}$ \Comment{\textcolor{magenta}{Ref-channel RIR parameter estimated}}

    \State $\hat{X} \gets \text{STFT}(\hat{x}_1^0)$
    
    \For{$c = 2, \dots, C$} \Comment{\textcolor{magenta}{Non-ref-channel RIR estimation}}
        \State $Y^c \gets \text{STFT}(\hat{y}_0)$
        \State $\hat{H}^c \gets \text{FCP}(Y^c, \hat{X})$ \Comment{\textcolor{magenta}{Non-ref-channel RIRs estimated}}
    \EndFor

    \State $\mathcal{L}_{\text{ref}}\gets \|S_{\text{comp}}({y_1})-S_{\text{comp}}(A_{\hat{\psi_1}}(\hat{x}_1^0))\|_2^2$
    \State $\mathcal{L}_{\text{non-ref}}\gets\sum_{c=2}^C \|S_{\text{comp}}({y_c})-S_{\text{comp}}(A_{\hat{H}^c}(\hat{x}_1^0))\|_2^2$
    
    \State $g_n \gets \zeta(\tau_n)\nabla_{{x}_n} (\mathcal{L}_{\text{ref}} + \lambda' \mathcal{L}_{\text{non-ref}})$ \Comment{\textcolor{magenta}{LH score approx.}}
    \State ${x}_1^{n-1} \gets {x}_1^n - \sigma_n (\sigma_{n-1} - \sigma_n)(s_n + g_n)$ \Comment{\textcolor{magenta}{Update step}}
\EndFor
\State \Return ${x}_1^0$ \Comment{\textcolor{magenta}{Reconstructed audio signal}}
\end{algorithmic}
\end{algorithm}

\section{Experiment}
\label{sec:experiment}

\subsection{Diffusion Model}

For the diffusion denoising architecture, we use the waveform domain U-Net
proposed in~\cite{dpssep}.
It has been open-sourced in \textit{audio-diffusion-pytorch/v0.1.3}\footnote{See \url{https://github.com/archinetai/audio-diffusion-pytorch/tree/v0.1.3}.}~\cite{mousai}. We find that this U-Net produces better performance and has faster sampling time than the modified NCSN++ model in BUDDy~\cite{buddy}. A direct comparison can be found in Table~\ref{tab:si_sdr_table} and Table~\ref{tab:time}. We train the diffusion model on the train-clean-\{$100$,$360$\} subsets of the LibriTTS dataset~\cite{libritts}. The subsets contain $\sim$$460$ hours of clean speech with more than $1,000$ speakers. The diffusion training follows the training recipe in BUDDy. For sampling, we use the diffusion scheduler and first-order sampler exactly the same as in BUDDy, with $200$ sampling steps. We encourage the readers to check our code in: \url{https://github.com/USDDPS/USDDPS_code}.

\subsection{Dataset and Metrics}
\label{ssec:dataset}

For evaluation, we use the WSJ0CAM-DEREVERB dataset, which has been used in USDnet~\cite{usdnet} and other supervised speech dereverberation studies~\cite{fcp1, tfgridnet}. It uses clean speech utterances from the WSJ0CAM corpus~\cite{wsjcam0}, to simulate $39,293$ ($\sim$$77.7$h), $2,968$ ($\sim$$5.6$h), and $3,262$ ($\sim$$6.4$h) mixtures for training, validation, and testing, respectively. Each mixture is synthesized by randomly sampling room acoustics and positions for the speaker and microphones. An $8$-channel circular microphone array (with a diameter of $20$ cm) records speech at distances between $0.75$ and $2.5$ m, with reverberation times (T60) between $0.2$ and $1.3$ s. Diffuse air-conditioning noise from the REVERB dataset \cite{REVERB} is added at randomly-chosen SNRs from $5$ to $25$ dB. The sampling rate is $16$ kHz. We use a subset of $100$ mixtures in the validation set to tune our hyperparameters for ablation studies, and use the full test set to evaluate and compare models.

For evaluation metrics, we use the direct-path signal at the reference channel (first microphone) as the reference signal for metric computation. We report perceptual evaluation of speech quality (PESQ) \cite{pesq} for perceptual quality,  extended short-time objective intelligibility (eSTOI) \cite{estoi} for speech intelligibility, and SI-SDR~\cite{sisdr} for sample-level consistency. We use the \textit{python-pesq} toolkit
to report narrow-band PESQ, and the \textit{pystoi} toolkit to report eSTOI. Note that since unsupervised methods like our USD-DPS, USDnet, and BUDDy often generate outputs misaligned with the ground-truth signal, resulting in low SI-SDR, which is sensitive to sample-level alignment.

\begin{table*}[t]
\vspace{-7pt}
\scriptsize
\renewcommand{\arraystretch}{1.1}  
\centering
\caption{Averaged SI-SDR (dB), PESQ and eSTOI results for $2$-, $4$-, and $8$-channel speech dereverberation on test set of WSJ0CAM-DEREVERB dataset. Bolded numbers indicate best
performance among unsupervised methods.}
\setlength\tabcolsep{4.6pt}
\label{tab:si_sdr_table}
\begin{tabular}{
    l  
    c  
    c  
    *{3}{S[round-precision=1,table-format=2.1]S[round-precision=2,table-format=1.2]S[round-precision=3,table-format=2.2]} 
}
\toprule
\multirow{2}{*}{Method} 
& \multirow{2}{*}{\textbf{Unsup.}} 
& \multirow{2}{*}{\makecell{\textbf{Multi} \\ \textbf{Channel}}}
& \multicolumn{3}{c}{\textbf{2 mics}} 
& \multicolumn{3}{c}{\textbf{4 mics}} 
& \multicolumn{3}{c}{\textbf{8 mics}} \\
\cmidrule(lr){4-6} \cmidrule(lr){7-9} \cmidrule(lr){10-12}
& & & {SI-SDR$\uparrow$} & {PESQ$\uparrow$} & {eSTOI$\uparrow$}
      & {SI-SDR$\uparrow$} & {PESQ$\uparrow$} & {eSTOI$\uparrow$}
      & {SI-SDR$\uparrow$} & {PESQ$\uparrow$} & {eSTOI$\uparrow$} \\
\midrule
Mixture &  &  & -3.6  & 1.64 & 0.494  & -3.6 & 1.64 & 0.494 & -3.6 & 1.64 & 0.494  \\
\midrule
\midrule		

WPE~\cite{wpe1}                    & \textcolor{green}{\cmark} & $\xmark$ & -3.1         & 1.67          & 0.512          & -3.1         & 1.67          & 0.512          & -3.1        & 1.67          & 0.512          \\
WPE~\cite{wpe1}                    & \textcolor{green}{\cmark} & $\cmark$ & -1.0         & 1.90          & 0.630          & -1.0         & 1.98          & 0.665          & -0.1        & 1.97          & 0.656          \\
USDnet~\cite{usdnet}               & \textcolor{green}{\cmark} & $\cmark$ & 2.1          & 2.37          & 0.745          & 2.3          & 2.53          & 0.751          & 2.7         & 2.45          & 0.761          \\
BUDDy (NCSN++)~\cite{buddy}        & \textcolor{green}{\cmark} & $\xmark$ & 1.0          & 2.31          & 0.778          & 1.0          & 2.31          & 0.778          & 1.0         & 2.31          & 0.778             \\
BUDDy (1D U-Net)                     & \textcolor{green}{\cmark} & $\xmark$ & 2.1          & 2.49          & 0.802          & 2.1          & 2.49          & 0.802          & 2.1          & 2.49          & 0.802  \\
\midrule
MC-FCP (proposed)              & \textcolor{green}{\cmark} & $\cmark$ & -13.8        & 1.63          & 0.479          & -13.8        & 1.64          & 0.487          & -11.7       & 1.70          & 0.516  \\
MC-BUDDy (proposed)            & \textcolor{green}{\cmark} & $\cmark$ & -3.3         & 2.67          & 0.834          & -5.5         & 2.74          & \bfseries 0.884          & -6.6        & 2.68          & 0.838  \\
USD-DPS (proposed) & \textcolor{green}{\cmark} & $\cmark$ 
& \bfseries 3.6 & \bfseries 2.79 & \bfseries 0.871 
& \bfseries 3.8 & \bfseries 2.90 & \bfseries 0.884 
& \bfseries 3.5 & \bfseries 2.94 & \bfseries 0.879 \\

\midrule
DNN-WPE~\cite{dnnwpe}               & \textcolor{red}{\xmark} & $\cmark$ & 1.6          & 2.00          & 0.697          & 2.9          & 2.08          & 0.731          & 2.9         & 2.09          & 0.728          \\
NB-LSTM~\cite{nbblstm}         & \textcolor{red}{\xmark} & $\cmark$ & 4.0 & 2.20 & 0.722 & 6.3 & 2.45 & 0.789 & 7.8 & 2.69 & 0.827 \\
NBC~\cite{quan2022multichannelspeechseparationnarrowband}         & \textcolor{red}{\xmark} & $\cmark$ & 7.6 & 2.63 & 0.824 & 11.1 & 3.08 & 0.898 & 13.0 & 3.31 & 0.926 \\
\bottomrule
\end{tabular}
\end{table*}

\subsection{Method Configurations}

{\bf MC-BUDDy}: For RIR modeling, we follow the setups in BUDDy, except that we set the number of frames of the RIR model $N_h$ to be $150$, corresponding to $\sim$$1.2$ seconds.  Each channel has its own independent RIR model, but only the reference channel's projection step contains the $\delta\oplus (\cdot)$ operation described in Sec. \ref{sec:likelihood}.
Following \cite{buddy, cqt}, $\zeta(\tau)$ in Eq.~\ref{eq:mcbuddy_likelihood_aprox} is set to $\zeta(\tau) = \frac{\zeta\sqrt{L}}{\tau\|G\|_2}$, where $\zeta$ is set to $0.8$ after careful tuning and $G$ is the gradient component of Eq.~\ref{eq:mcbuddy_likelihood_aprox}:
{\small
\begin{equation}\label{eq:G}
    G= \nabla_{x_1^\tau}\sum_{c=1}^{C}\left\|S_{\text{comp}}(y_c)-S_{\text{comp}}(A_{\hat{\psi}_c}(\hat{x}_1^0))\right\|_2^2
\end{equation}
}

{\bf USD-DPS}: USD-DPS's reference-channel RIR model follows MC-BUDDy above. For FCP used for non-reference RIR estimation, we use STFT with $32$ms FFT size, $8$ms hop size, and square root Hanning window. We set the FCP filter length $N_h'$ to $60$. We set $\epsilon$ in Eq.~\ref{eq:fcp_lambda} to $0.001$. We also set $\zeta(\tau) = \frac{\zeta\sqrt{L}}{\tau\|G\|_2}$ as in Eq.~\ref{eq:usddps_likelihood_aprox}, where $G$ is the gradient term in Eq.~\ref{eq:usddps_likelihood_aprox}. Again, through careful tuning, we set $\zeta=0.8$ and $\lambda=0.6$ as in Eq.~\ref{eq:usddps_likelihood_aprox}.


\subsection{Baselines}
We consider WPE \cite{wpe1} and USDnet \cite{usdnet} as unsupervised baselines, and DNN-WPE \cite{dnnwpe}, NB-BLSTM~\cite{nbblstm} and NBC~\cite{quan2022multichannelspeechseparationnarrowband} as supervised baselines. We also develop an MC-FCP baseline for ablation, where we use FCP for all channels' RIRs estimation, without using the RIR model at all. For WPE, we use the implementation in the \textit{torchiva} toolkit~\cite{torchiva}. The STFT configuration of WPE is the same as the setting in the sub-band filtering operator in our experiment. The filter tap is tuned to $37$ in monaural cases, to $20$ in $2$-channel cases, $10$ in $4$-channel cases, and $5$ in $8$-channel cases. The prediction delay is $3$ frames, and $3$ iterations are performed. For the USDnet baseline, we follow the result in TABLE VII's row 2a, 2b, and TABLE VIII's row 3b in \cite{usdnet}. For the supervised baselines, we use the default architecture of NB-BLSTM and NBC. For DNN-WPE, we use NBC as the DNN, and the STFT configuration follows the WPE baseline. We also use use single-channel BUDDy as a baseline, where we have one version that uses the NCSN++ U-Net as in~\cite{buddy}, and another version uses our 1-D U-Net for fair comparison.


\section{Results and Analysis}
\label{sec:results}

Table \ref{tab:si_sdr_table} reports the dereverberation results
on WSJ0CAM-DEREVERB, and Table~\ref{tab:time} shows the average inference time for processing an $8$-channel mixture. For $2$-channel dereverberation, we use microphone $1$ and $5$; for $4$-channel dereverberation, we use microphone $1$, $3$, $5$ and $7$; and for $8$-channel dereverberation, all the $8$ microphones are used. First, we compare single-channel BUDDy with different diffusion architectures.
\begin{table}[ht]
\scriptsize
\centering
\caption{Averaged inference time per mixture in seconds when evaluating on $8$-channel WSJ0CAM-DEREVERB test set. The average duration of each mixture is 7.0 seconds.}
\label{tab:time} 
\begin{tabular}{lccccc}
\toprule
\textbf{Method} & \textbf{BUDDy} & \textbf{BUDDy} & \textbf{MC-BUDDy} & \textbf{USD-DPS} \\
\midrule
Architecture                           & NCSN++   & 1D U-Net   &    1D U-Net  &    1D U-Net  \\
Processing Time (s)                & 119.67   & 64.78     & 387.75      & 114.38           \\
PESQ                           &  2.31    &  2.49     & 2.68        & 2.94            \\
\bottomrule
\vspace{-15pt}
\end{tabular}
\end{table}

It is clear that using a 1D U-Net not only performs better than NCSN++ in all the metrics, but also has faster inference time as shown in Table~\ref{tab:time}. Thus, for MC-BUDDy and USD-DPS, we choose to use the 1D U-Net. Then, from Table~\ref{tab:si_sdr_table}, we find that USD-DPS and MC-BUDDy perform much better than USDnet, suggesting the benefits of using a diffusion speech prior. Comparing MC-BUDDy and BUDDy, we observe clear gains in PESQ and eSTOI by leveraging multi-channel measurements. However, MC-BUDDy shows negative SI-SDR, which is probably because it uses independent RIR models for all channels. In reality, different channels' RIR should have some similar properties. By comparing MC-BUDDy with USD-DPS, which only uses an RIR model for the reference channel, we find that USD-DPS performs better in all the metrics, and is also much more efficient, as shown in Table~\ref{tab:time}. Lastly, we observe that MC-FCP does not work, which means that only using FCP for all channels' RIRs is not feasible. This observation aligns with the USDnet's result when using complex spectral mapping~\cite{usdnet}.

We then compare USD-DPS with supervised baselines as in Table~\ref{tab:si_sdr_table}. First, USD-DPS outperforms DNN-WPE in all metrics. Compared with NB-LSTM, USD-DPS is much better for PESQ and eSTOI, but worse in terms of SI-SDR, possibly because SI-SDR is sensitive to misalignment. When compared with NBC, a much stronger supervised baseline, USD-DPS only performs better in PESQ and eSTOI in the 2-channel scenario, but performs worse in all the other cases.
This is possibly because supervised models can learn the noise distribution from the training set to have good denoising performance, while USD-DPS simply assumes white noise and shows relatively weak denoising performance. One possible solution is to use more complicated noise modeling (e.g., use another noise diffusion model), which we leave to future research. Also, NBC and NB-LSTM in Table~\ref{tab:si_sdr_table} need to be trained for different settings, while USD-DPS is training-free for any settings. 
\section{Conclusion}
We have proposed USD-DPS, an unsupervised, generative method for multi-channel blind speech dereverberation. USD-DPS uses a clean speech diffusion prior and a novel likelihood approximation to enable posterior sampling. For the likelihood approximation, we propose to use an RIR model for reference-channel RIR estimation and use FCP for non-reference channels' RIRs estimation. We find that this combination can balance RIR modeling and RIR estimation efficiency, yielding better dereverberation than all existing unsupervised methods. Moving forward, we will explore USD-DPS for more general array inverse problems like simultaneous speech enhancement, separation, and dereverberation, with a microphone array.

\clearpage
\bibliographystyle{IEEEtran}
\bibliography{refs25}








\end{document}